\documentclass[letterpaper, 10 pt, conference]{ieeeconf}  % Comment this line out if you need a4paper

\IEEEoverridecommandlockouts                              % This command is only needed if 
\newif\iflong
\longtrue

\title{\LARGE \bf
Sign-Perturbed Sums (SPS) with Instrumental Variables\\ for the Identification of ARX Systems \iflong -- Extended Version\fi
}
\author{Valerio Volpe$^{\dag}$\thanks{The work of B.\ Cs.\ Cs\'aji was supported by the Hungarian Scientific Research Fund (OTKA), pr.~no.~113038, and by the J\'anos Bolyai Research Fellowship of the Hungarian Academy of Sciences, pr.~no.~BO/00683/12/6. The work of A.\ Car\`e and E.\ Weyer was supported by the Australian Research Council (ARC) under Discovery Grant DP130104028. The work of M.\ C.\ Campi was partly supported by MIUR - Ministero dell'Istruzione, dell'Universit\`a e della Ricerca.}
\thanks{V.\ Volpe$^{\dag}$ and M.\ C.\ Campi$^{\ddag}$ are with Department of Information Engineering, University of Brescia, Via
Branze 38, 25123 Brescia, Italy; {\tt\footnotesize (email: v.volpe@studenti.unibs.it,} {\tt\footnotesize marco.campi@unibs.it)}}\and Bal\'azs Cs.\ Cs\'aji$^{\S}$\thanks{B.\ Cs.\ Cs\'aji$^{\S}$ is with Fraunhofer Project Center at the Institute for Computer Science and Control (SZTAKI), Hungarian
Academy of Sciences (MTA), Kende utca 13--17, Budapest, Hungary, H-1111; {\tt\footnotesize (email: balazs.csaji@sztaki.mta.hu)}
}\and Algo Car\`e$^{*}$\thanks{A.\ Car\`e$^{*}$ and E.\ Weyer$^{**}$ are with Department of Electrical and Electronic Engineering, Melbourne School
of Engineering, The University of Melbourne, 240 Grattan Street, Parkville,
Melbourne, Victoria, 3010, Australia; {\tt\footnotesize (email: $\{$algo.care,ewey$\}$@unimelb.edu.au)}}\and Erik Weyer\thanks{}$^{**}$\and Marco C. Campi$^{\ddag}$\thanks{}% <-this % stops a space
}
\usepackage{amsmath}
\usepackage{amssymb}
\usepackage{mathrsfs}
\usepackage{hyperref}
\usepackage{graphicx}
\usepackage{float}
\makeatletter
\newcommand{\labitem}[2]{%
\def\@itemlabel{\textbf{#1}}
\item
\def\@currentlabel{#1}\label{#2}}
\makeatother

\newlength{\dhatheight}
\newcommand{\doublehat}[1]{%
    \settoheight{\dhatheight}{\ensuremath{\widehat{#1}}}%
    \addtolength{\dhatheight}{-0.35ex}%
    \widehat{\vphantom{\rule{1pt}{\dhatheight}}%
    \smash{\widehat{#1}}}}

\newtheorem{lemma}{Lemma}

\newtheorem{definition}{Definition}
\newtheorem{theorem}{Theorem}

\begin{document}

\maketitle
\thispagestyle{empty}
\pagestyle{empty}

%%%%%%%%%%%%%%%%%%%%%%%%%%%%%%%%%%%%%%%%%%%%%%%%%%%%%%%%%%%%%%%%%%%%%%%%%%%%%%%%
\begin{abstract}

We propose a generalization of the recently developed system identification method called Sign-Perturbed Sums (SPS). The proposed construction is based on the instrumental variables estimate and, unlike the original SPS, it can construct non-asymptotic confidence regions for linear regression models where the regressors contain past values of the output. Hence, it is applicable to ARX systems, as well as systems with feedback. We show that this approach provides regions with exact confidence under weak assumptions, i.e., the true parameter is included in the regions with a (user-chosen) exact probability for any finite sample. The paper also proves the strong consistency of the method and proposes a computationally efficient generalization of the previously proposed ellipsoidal outer-approximation. Finally, the new method is demonstrated through numerical experiments, using both real-world and simulated data.
\end{abstract}

%%%%%%%%%%%%%%%%%%%%%%%%%%%%%%%%%%%%%%%%%%%%%%%%%%%%%%%%%%%%%%%%%%%%%%%%%%%%%%%%
\section{Introduction}

Estimating parameters of partially unknown systems based on observations corrupted by noise is a classic problem in signal processing, system identification, machine learning and statistics \cite{Gevers2006,Ljung1999,Ljung2010,Ljung1994,Soderstrom1989}. Many standard methods are available which perform point estimations. Given an estimate, it is an intrinsic task to evaluate how close the estimated parameter is to the true one and such evaluation often comes in the form of a confidence region. Confidence regions are especially important for problems where the quality, stability or safety of a process has to be guaranteed.

The Sign-Perturbed Sums (SPS) method was presented in \cite{SPSPaper1,SPSPaper2,WCC2013,kolumban2015perturbed}. Implementations of the method based on interval analysis have been proposed in \cite{kieffer2013guaranteed,kieffer2013guaranteedIFAC, kieffer2014guaranteed}, and an application of the method under a different set of assumptions has been presented in \cite{senov2014exact}. The main feature of the SPS method is that it constructs confidence regions which have an exact probability of containing the system's true parameter based on a finite number of observed data.

The SPS method of \cite{SPSPaper2} and \cite{WCC2013} provides exact confidence regions for the true parameter only when the regressors are exogenous (i.e., they do not depend on the noise terms), which is not the case with ARX systems, or, e.g., when feedback is involved. Generalizing the method to the case where the regressors can depend on the noise terms is of high practical importance.

In \cite{SPSPaper1} an SPS method which deals with ARX systems has been given, and even more general systems have been considered in \cite{kolumban2015perturbed, CDC2012}. However, these extensions introduce complications in the simple algorithm of \cite{SPSPaper2} and \cite{WCC2013}, which make the method more challenging to analyze and more difficult to implement and run. In this paper we follow an alternative path, and show that an instrumental variables approach allows for notable simplifications in the algorithms. This leads, on the one hand, to computationally tractable methods for building regions and, on the other hand, to easy-to-prove, and quite general, strong consistency results.

The paper is organized as follows. In the next section we state the problem setting and our main assumptions. Then, the generalization of the SPS algorithm is presented in Section \ref{SPSIV}, and in Section \ref{TheoreticalResults} we illustrate the theoretical properties of the constructed confidence regions. Subsequently, we give a simplified construction by way of an outer ellipsoidal approximation algorithm similar to that developed in \cite{SPSPaper2} for the case of exogenous regressors. Finally, in Section \ref{NumericalExp}, we show two applications of the generalized SPS algorithm with numerical experiments, using both real-world and computer generated data. The proofs can be found in the \iflong appendices.\else extended version of this paper, \cite{extendedPaper}.\fi
\section{Problem setting}
This section presents the linear regression problem and introduces our main assumptions.

\subsection{Data generation}
The data are generated by the following system
\begin{equation}
\label{LinearRegression}
Y_t \, \triangleq \,
\varphi_t^\mathrm{T} \theta^* + N_t,
\end{equation}
where $Y_t$ is the output, $N_t$ is the noise, $\varphi_t$ is the regressors, and $t$ is the discrete time index. Parameter $\theta^*$ is the true parameter to be estimated. The random variables $Y_t$ and $N_t$ are real-valued, while $\varphi_t$ and $\theta^*$ are  $d$-dimensional real vectors. We consider a finite sample of size $n$ which consists of the regressors $\varphi_1, \dots, \varphi_n$ and the outputs $Y_1, \dots, Y_n$.

In addition, we assume that a set of instrumental variables $\{\psi_t\}_{t=1}^n$ is available to the user. The terms in the sequence must be correlated with the data and independent of the noise. Typically, past or filtered past inputs are used as instrumental variables.
\subsection{Examples}
There are many examples in signal processing and control of systems taking the form of (\ref{LinearRegression}), see \cite{Ljung1999,Soderstrom1989}. 
An important example is the widely used ARX model
\begin{equation*}
Y_t = \sum_{i=1}^{d_1}a_i^*Y_{t-i} + \sum_{i=1}^{d_2}b_i^*U_{t-i} + N_t
\end{equation*}
where $\varphi_t\!=\![Y_{t-1},\ldots,Y_{t-d_1},U_{t-1},\ldots,U_{t-d_2}]^\mathrm{T}\!$ consists of past outputs and inputs, and the true parameter $\theta^*\in \mathbb{R}^{d_1+d_2}$  is the vector $[a_1^*, \ldots,a_{d_1}^*,b_1^*,\ldots,b_{d_2}^*]^\mathrm{T}$. An instrumental variables sequence $\{\psi_t\}$ can be easily obtained from the data. In particular, the instrumental variables vector can be constructed from the regressor $\varphi_t$ by replacing the (noise-dependent) outputs with some other variables, such as delayed inputs, or noise-free reconstructed output terms, that can be computed using a guess of the true system parameter. The latter approach, in particular, is used and showed in Section \ref{NumericalExp}.
\subsection{Basic assumptions}
Our assumptions on the regressors, the instrumental variables and the noise are:

\begin{enumerate}
\labitem{A1}{AssumpNoise}$\{N_t\}$ is a sequence of independent random variables. Each $N_t$ has a symmetric probability distribution about zero.
\smallskip
\labitem{A2}{Invertibility}$\det(V_n)\neq 0$ almost surely, where $$V_n \triangleq \frac{1}{n}\sum_{t=1}^n\psi_t \varphi_t^\mathrm{T}.$$
\end{enumerate}
Note that \ref{Invertibility} implies that matrix $H_n\triangleq\frac{1}{n}\sum_{t=1}^{n}\psi_t\psi_t^\mathrm{T}$ is (almost surely) invertible.

Like the SPS of \cite{SPSPaper2} the assumptions are rather mild, since there are no moment or density requirements on the noise terms, and their distributions can change with time and need not be known. The strongest assumption on the noise is that it forms an independent sequence, but it can be somehow relaxed with the suitably modified Block SPS \cite{SPSPaper2}. The core assumption is the symmetricity of the noise. Many standard distributions satisfy this property. These weak requirements make the method widely applicable.
\section{Sign-Perturbed Sums with instrumental variables}\label{SPSIV}
In this section we introduce the generalization of SPS using instrumental variables.
\subsection{Intuitive idea}
First, recall that the instrumental variables estimate $\hat{\theta}_n$ comes as the solution to a modified version of the normal equations, i.e.,
\begin{equation}
\label{EquationIV}
\sum_{t=1}^{n}\psi_t(Y_t - \varphi_t^\mathrm{T}\theta)=0,
\end{equation}
and the instrumental variables (IV) estimate is
\begin{equation*}
\hat{\theta}_n\triangleq\left(\sum_{t=1}^{n}\psi_t\varphi_t^\mathrm{T}\right)^{-1}\sum_{t=1}^{n}\psi_tY_t.
\end{equation*}

Then, referring to the same ideas as in \cite{SPSPaper2} for the construction of the SPS method, we can build $m-1$ sign-perturbed versions of equation \eqref{EquationIV}, and define the {sign\em -perturbed sums} as
\[
S_i(\theta)\triangleq H_n^{-\frac{1}{2}}\frac{1}{n}\sum_{t=1}^{n}\psi_t\alpha_{i,t}(Y_t-\varphi_t^\mathrm{T}\theta),
\]
$i \in \{1, \ldots,m-1\}$, where $H_n^{1/2}$ is the principal square root of $H_n$, which is introduced in order to give a better shape to the confidence regions, and $\{\alpha_{i,t}\}$ are
i.i.d.\ Rademacher variables,  i.e., they take on the values $\pm1$ with probability 1/2 each. Also, without applying sign-perturbations, we can define the {\em reference sum} as
\[
S_0(\theta)\triangleq H_n^{-\frac{1}{2}}\frac{1}{n}\sum_{t=1}^{n}\psi_t(Y_t-\varphi_t^\mathrm{T}\theta).
\]

An important property of these functions is that corresponding to $\theta=\theta^*$ we have
\begin{equation*}
S_0(\theta^*)=H_n^{-\frac{1}{2}}\frac{1}{n}\sum_{t=1}^{n}\psi_tN_t,
\end{equation*}
\begin{equation*}
S_i(\theta^*)=H_n^{-\frac{1}{2}}\frac{1}{n}\sum_{t=1}^{n}\alpha_{i,t}\psi_tN_t=H_n^{-\frac{1}{2}}\frac{1}{n}\sum_{t=1}^{n}\pm\psi_tN_t,
\end{equation*}
and such variables are uniformly ordered, i.e., once the values of $\{\|S_i(\theta^*)\|^2\}_{i=0}^{m-1}$have been sorted according to a particular strict total order, any $\|S_i(\theta^*)\|^2$ has the same probability of being ranked in a given position\iflong\:(see Appendix A)\else\:(see \cite[Appendix A]{extendedPaper})\fi. This observation is crucial to SPS since it builds the confidence regions by excluding those $\theta$ for which $\|S_0(\theta)\|^2$ is among the $q$ largest ones, and the  so constructed confidence set has exact probability $1-q/m$ of containing the true parameter\footnote{Notice that many $q$ and $m$ pairs give the same ratio $q/m$. Refer to \cite{SPSPaper2} for more discussion on the choice of $q$ and $m$.}.

Moreover, when $\|\theta'-\theta^*\|$ is large $\|S_0(\theta')\|^2$ tends to be the largest of the $m$ functions. Therefore, defining $\pi$ as a {\em random} permutation of the set $\{0,\ldots,m-1\}$ and the strict total order by\footnote{The random permutation $\pi$ is used to break ties in case two
different $\|S_i(\theta')\|^2$
variables take on the same value.}
\[Z_j \succ_{\pi} Z_k \Leftrightarrow\left(Z_j > Z_k\right) \vee \left(Z_j = Z_k \wedge\pi(j) > \pi(k)\right),
\]
where $Z_i=\|S_i(\theta')\|^2$, it happens that values far away from $\theta^*$ are excluded from the confidence set.
\subsection{Formal construction of the confidence region}
The pseudocode of the generalized SPS algorithm is presented in two parts. The initialization (Table \ref{InitTab}) sets the main global parameters and generates the random objects needed for the construction. In the initialization, the user provides the desired confidence probability $p$. The second part (Table \ref{IndTab}) evaluates an indicator function, SPS-Indicator$(\theta)$, which determines if a particular parameter $\theta$ is included in the confidence region.
{\renewcommand{\arraystretch}{1.3}
\begin{table}[H]
\normalsize
\begin{center}
\begin{tabular}{|rlll|}
\hline
\multicolumn{4}{|c|}{\scshape Pseudocode: SPS-Initialization} \\
\hline \hline 1. & \multicolumn{3}{l|}{Given a (rational) confidence probability $p \in (0,1)$,} \\
 & \multicolumn{3}{l|}{set integers $m > q > 0$ such that $p = 1 - q/m$;}\\
2. & \multicolumn{3}{l|}{Calculate the outer product} \\
& \multicolumn{3}{c|}{$H_n\, \triangleq\, \frac{1}{n}\sum\limits_{t=1}^n\psi_t \psi_t^\mathrm{T}$,}\\
& \multicolumn{3}{l|}{ and find the principal square root $H_n^{1/2}$, such that} \\
& \multicolumn{3}{c|}{$H_n^{1/2}H_n^{1/2}=H_n$;}\\
3. & \multicolumn{3}{l|}{Generate $n\,(m-1)$ i.i.d.\ random signs $\{\alpha_{i,t}\}$ with} \\
& \multicolumn{3}{c|}{$\mathbb{P}(\alpha_{i,t} = 1)\, = \,\mathbb{P}(\alpha_{i,t} = -1) \,=\, \frac{1}{2}$,}\\
& \multicolumn{3}{l|}{for $i \in \{1, \dots, m-1\}$ and $t \in \{1, \dots, n\}$;}\\
4. & \multicolumn{3}{l|}{Generate a random permutation $\pi$ of the set}\\
& \multicolumn{3}{l|}{$\{0, \dots, m-1\}$, where each of the $m!$ possible}\\
& \multicolumn{3}{l|}{permutations has the same probability $1/(m!)$}\\
& \multicolumn{3}{l|}{to be selected.}\\
\hline
\end{tabular}
\end{center}
\caption{}\label{InitTab}
\vspace{-5mm}
\end{table}}
{\renewcommand{\arraystretch}{1.3}
\begin{table}[H]
\normalsize
\begin{center}
\begin{tabular}{|rlll|}
\hline
\multicolumn{4}{|c|}{\scshape Pseudocode: SPS-Indicator\,(\,$\theta$\,)}\\
\hline \hline 1. & \multicolumn{3}{l|}{For the given $\theta$, compute the prediction errors }\\
& \multicolumn{3}{l|} {for $t \in \{1, \dots, n\}$} \\
 & \multicolumn{3}{c|}{${\varepsilon}_t(\theta)\,\triangleq\, Y_t - \varphi_t^\mathrm{T}\theta$;}\\
2. & \multicolumn{3}{l|}{Evaluate} \\
& \multicolumn{3}{l|}{\hspace{15mm}$S_0(\theta) \triangleq H_n^{-\frac{1}{2}} \frac{1}{n}\sum\limits_{t=1}^{n}{\, \psi_t {\varepsilon}_t(\theta)}$,}\\
& \multicolumn{3}{l|}{\hspace{15mm}$ S_i(\theta) \triangleq H_n^{-\frac{1}{2}} \frac{1}{n}\sum\limits_{t=1}^{n}{\, \alpha_{i,t} \, \psi_{t}{\varepsilon}_t(\theta)}$,}\\
& \multicolumn{3}{l|}{for $i \in \{1, \dots, m-1 \}$;}\\
3. & \multicolumn{3}{l|}{Order scalars $\{\|S_i(\theta)\|^2\}$ according to $\succ_{\pi}$;}\\
4. & \multicolumn{3}{l|}{Compute the rank $\mathcal{R}(\theta)$ of $\|S_0(\theta)\|^2$ in the ordering} \\
& \multicolumn{3}{l|} {where $\mathcal{R}(\theta) = 1$ if $\|S_0(\theta)\|^2$ is the smallest in the} \\
& \multicolumn{3}{l|} {ordering, $\mathcal{R}(\theta) = 2$ if $\|S_0(\theta)\|^2$ is the second small-} \\
& \multicolumn{3}{l|} {est, and so on;}\\
6. & \multicolumn{3}{l|}{Return $1$ if $\mathcal{R}(\theta) \leq m-q$, otherwise return $0$.}\\
\hline
\end{tabular}
\end{center}
\caption{}\label{IndTab}
\vspace{-5mm}
\end{table}}

Using this construction, we can define the $p$-level SPS {\em confidence region} as follows
\begin{equation*}
\widehat{\Theta}_n \triangleq \left\{ \theta \in \mathbb{R}^d \!:\! \text{SPS-Indicator}(\theta) = 1\right\}.
\end{equation*}

Note that, corresponding to the instrumental variables estimate
$\hat{\theta}_n$, it holds that $S_0(\hat{\theta}_n)=0$. Therefore, with exception of pathological cases, $\hat{\theta}_n$  is included in the SPS confidence region, and the set is built around $\hat{\theta}_n$.
\section{Theoretical results}\label{TheoreticalResults}
\subsection{Exact confidence}
The most important property of the SPS method is that the generated regions have {\em exact} confidence probabilities for any {\em finite} sample. The following theorem holds.
\medskip
\begin{theorem}\label{TheoremExact}
{\em Assuming \ref{AssumpNoise} and \ref{Invertibility}, the confidence probability of the constructed confidence region is exactly $p$, that is,}
\begin{equation*}
\mathbb{P}\big(\theta^* \in \widehat{\Theta}_n\big)\, =\, 1 - \frac{q}{m} \, = \, p.
\end{equation*}
\end{theorem}
\medskip

The proof of the theorem, which is along the lines of the proof of Theorem 1 of \cite{SPSPaper2}, can be found in \iflong{Appendix A}\else\cite{extendedPaper}\fi.
Since the confidence probability is exact, no conservatism is introduced. Moreover, the statistical assumptions imposed on the noise are rather weak. Indeed the noise distribution can change during time, and there are no moment or density requirements whatsoever.
\subsection{Strong consistency}
An important aspect of the confidence region is its size. Clearly for any finite sample the size of the region depends much on the statistical properties of the noise. However, we show that asymptotically the SPS regions become smaller and smaller, shrinking to the true parameter. Indeed the SPS algorithm is {\em strongly consistent}, under the following (rather mild) assumptions.
\begin{enumerate}
\labitem{A3}{RPosDef}There exists a positive definite matrix $H$ such that
\[
\lim\limits_{n \rightarrow \infty}H_n = H,\, \mbox{almost surely}.
\]
\labitem{A4}{VInvertibility}There exists an invertible matrix $V$ such that
\[
\lim\limits_{n \rightarrow \infty}V_n = V,\, \mbox{almost surely}.
\]
\labitem{A5}{RegressorGrowthRestriction}({\em regressor growth rate restriction}):
\[
 \sum_{t=1}^{\infty}\frac{\lVert \varphi_t\rVert^4}{t^2} < \infty, \, \mbox{almost surely}.
 \]
 \labitem{A6}{InstrumentsGrowthRestriction}({\em instruments growth rate restriction}):
 \[
  \sum_{t=1}^{\infty}\frac{\lVert \psi_t\rVert^4}{t^2} < \infty,\, \mbox{almost surely}.
  \]
 \labitem{A7}{NoiseGrowthRestriction}({\em noise variance growth rate restriction}):
 \[
  \sum_{t=1}^{\infty}\frac{\mathbb{E}[N_t^2]^2}{t^2} < \infty.
  \]
\end{enumerate}
The following theorem holds.
\smallskip
\begin{theorem}\label{TheoremStrongCons}Assuming \ref{AssumpNoise}, \ref{Invertibility}, \ref{RPosDef}, \ref{VInvertibility}, \ref{RegressorGrowthRestriction},
\ref{InstrumentsGrowthRestriction} and \ref{NoiseGrowthRestriction}, $\forall \varepsilon > 0$ there  almost surely exists an $N$ such that $\forall n > N, \hat{\Theta}_n\subseteq\{ \theta \in \mathbb{R}^d: \|\theta - \theta^*\|\leq\varepsilon\}$.
\end{theorem}
\medskip

The proof of the theorem can be found in \iflong{Appendix B}\else\cite{extendedPaper}\fi. The claim states that the confidence regions $\{\hat{\Theta}_n\}$ will eventually be included (almost surely) in any norm-ball centered at $\theta^*$ as the sample size increases. Although the regions generated by the generalization of SPS introduced in this paper have no theoretical guarantee of being bounded, they normally are, and, moreover, the strong consistency result implies that they are bounded with probability 1 asymptotically.
\section{Ellipsoidal approximation algorithm}
The purpose of the SPS-Indicator function is to check whether a given $\theta$ belongs to the confidence region or not. In particular, it 
computes the $\{\|S_i(\theta)\|^2\}_{i=0}^{m-1}$ functions for that specific $\theta$ and compares them.
This way the SPS region can be constructed by decomposing the space of interest in a grid, possibly very dense, and checking whether the points in the grid belongs to the region. However, this approach is {\em computationally demanding}, and it gets slower and slower as the dimensions increase. Here, we introduce a generalization of the ellipsoidal outer approximation algorithm previously introduced for the SPS of \cite{SPSPaper2, WCC2013}. The algorithm leads to an ellipsoidal over-bound that can be efficiently computed in polynomial time.
\iflong
\subsection{Ellipsoidal outer approximation}
Expanding $\|S_0(\theta)\|^2$ we find that it can be written as
\begin{align*}
\|S_0(\theta)\|^2\! &=\!\bigg[\frac{1}{n}\!\sum_{t=1}^n\psi_t(Y_t\!-\!\varphi_t^\mathrm{T}\theta)\bigg]^{\mathrm{T}}\!\!\!\!H_n^{-1}
\!\bigg[\frac{1}{n}\!\sum_{t=1}^n\psi_t(Y_t\!-\!\varphi_t^\mathrm{T}\theta)\bigg] \\ & =\!\bigg[\frac{1}{n}\!\sum_{t=1}^n\psi_t\varphi_t^\mathrm{T}(\theta\!-\!\hat{\theta}_n)\bigg]^\mathrm{T}\!\!\!\!H_n^{-1} \!\bigg[\frac{1}{n}\!\sum_{t=1}^n\psi_t\varphi_t^\mathrm{T}(\theta\!-\!\hat{\theta}_n)\bigg] \\ &
=\! (\theta\!-\!\hat{\theta}_n)^{\mathrm{T}}V_n^\mathrm{T}H_n^{-1}V_n(\theta\!-\!\hat{\theta}_n).
\end{align*}
Then, since we are looking for an ellipsoidal over-bound, we can ignore the random ordering used when $\|S_0(\theta)\|^2$ and $\|S_i(\theta)\|^2$ take on the same value, and just consider the set given by those values of $\theta$  at which $q$ of the $\|S_i(\theta)\|^2$ are larger {\em or equal} to $\|S_0(\theta)\|^2$, i.e.\
\begin{equation*}
\widehat{\Theta}_n \subseteq \left\{\theta \in \mathbb{R}^d\!:\!(\theta-\hat{\theta}_n)^\mathrm{T}V_n^\mathrm{T}H_n^{-1}V_n(\theta-\hat{\theta}_n)\leq r(\theta) \right\},
\end{equation*}
where $r(\theta)$ is the $q$\hspace{0.3mm}th largest value of functions $\{\|S_i(\theta)\|^2\}$, $i=1,\ldots,m-1$.

The idea is to find an over-bound by replacing $r(\theta)$ with a parameter independent $r$, thus obtaining an outer approximation that is a guaranteed confidence region for finitely many data points. Moreover, since it is described in terms of $\hat{\theta}_n, V_n, H_n$ and $r$, it comes with a compact representation.

\subsection{Convex programming formulation}
Comparing $\|S_0(\theta)\|^2$ with one single $\|S_i(\theta)\|^2$ function, we have
\begin{eqnarray*}
\lefteqn{
\{\,
\theta: \|S_0(\theta)\|^2 \leq \|S_i(\theta)\|^2
\,\}
} \\
& & \subseteq
\{\,
\theta: \|S_0(\theta)\|^2 \leq \sup_{\theta: \|S_0(\theta)\|^2 \leq \|S_i(\theta)\|^2}\|S_i(\theta)\|^2
\,\}. \nonumber
\end{eqnarray*}
The inequality $\|S_0(\theta)\|^2 \leq \|S_i(\theta)\|^2$ can be rewritten as
\begin{eqnarray*}
\lefteqn{(\theta-\hat{\theta}_{n})^\mathrm{T}V_n^\mathrm{T}H_n^{-1}V_n(\theta-\hat{\theta}_{n}) \leq}\\
& &
\theta^\mathrm{T}Q_i^\mathrm{T}H_n^{-1}Q_i\theta-2\theta^\mathrm{T}Q_i^\mathrm{T}H^{-1}_n{\rho}_{i}+\rho^{\mathrm{T}}_{i}H^{-1}_n{\rho}_{i},
\end{eqnarray*}
where matrix $Q_i$ and vector $\rho_i$ are defined as
\begin{align*}
Q_i&\triangleq\frac{1}{n}\sum_{t=1}^n\alpha_{i,t}\psi_t\varphi^\mathrm{T}_t,\\
\rho_{i} &\triangleq\frac{1}{n}\sum_{t=1}^n\alpha_{i,t} \psi_t Y_t.
\end{align*}
First, observe that it holds that
$$
\sup_{\theta: \|S_0(\theta)\|^2 \leq \|S_i(\theta)\|^2}\|S_i(\theta)\|^2=\sup_{\theta: \|S_0(\theta)\|^2 \leq \|S_i(\theta)\|^2}\|S_0(\theta)\|^2.
$$
Such supremum is finite only if the matrix $V_n^\mathrm{T}H_n^{-1}V_n - Q_i^\mathrm{T}H_n^{-1}Q_i$ is positive semidefinite. If this is the case, we want to compute such maximum. Thus, defining $z\triangleq H_n^{-\frac{1}{2}}V_n(\theta-\hat{\theta}_n)$, we can find the quantity $$\max_{\theta: \|S_0(\theta)\|^2 \leq \|S_i(\theta)\|^2}\|S_i(\theta)\|^2,$$
as the solution of the following quadratic programming problem with only one quadratic constraint
\begin{eqnarray}
&  \mathrm{maximize} & \|z\|^2 \nonumber \\
& \mbox{subject to } & z^\mathrm{T}A_{i}z+2 z^\mathrm{T}b_{i}+c_{i}\leq 0, \nonumber
\end{eqnarray}
where $A_{i}$, $b_{i}$ and $c_{i}$ are defined as
\begin{align*}
A_{i}&\triangleq I-H_n^{\frac{1}{2}\mathrm{T}}V_n^{\mathrm{-T}}Q_i^\mathrm{T}H_n^{-1}Q_iV_n^{-1}H_n^{\frac{1}{2}},\\
b_{i}&\triangleq H_n^{\frac{1}{2}\mathrm{T}}V_n^{\mathrm{-T}}Q_i^\mathrm{T} H_n^{-1}(\rho_i-Q_i\hat{\theta}_n),\\
c_{i}&\triangleq -\rho^\mathrm{T}_i H_n^{-1}\rho_i+2\hat{\theta}_n^\mathrm{T} Q_i^\mathrm{T} H_n^{-1}\rho_i- \hat{\theta}_n^\mathrm{T} Q_i^\mathrm{T} H_n^{-1} Q_i \hat{\theta}_n.
\end{align*}
This program is not convex in general, due to the fact that the Hessian of the quadratic constraint is not necessarily positive definite. However, it can be shown, \cite[Appendix B]{Boyd2009}, that {\em strong duality} holds, so that the value of the above optimization problem is equal to the value of its dual, which can be formulated as the following semi-definite programming problem
\begin{eqnarray}
\label{OptDual}
&  \mathrm{minimize} & \gamma \nonumber \\
& \mbox{subject to } &
\lambda\geq 0 \nonumber \\
& &
\left[\begin{array}{cc} -I+\lambda A_{i} & \lambda b_{i} \\ \lambda b_{i}^\mathrm{T} & \lambda c_{i} + \gamma\end{array}\right]\succeq 0,
\end{eqnarray}
where ``$\succeq 0$'' denotes that a matrix is positive semidefinite. This program is convex, and can be easily solved in polynomial time using, e.g., MATLAB and a toolbox such as CVX \cite{grant2008cvx}.

Defining $\gamma_i^*$ as the value of program \eqref{OptDual}, we have
$$
\{
\theta: \|S_0(\theta)\|^2 \leq \|S_i(\theta)\|^2
\}
\subseteq
\{
\theta: \|S_0(\theta)\|^2 \leq \gamma_i^*
\}.
$$
Thus,
\begin{equation*}
\widehat{\Theta}_n \subseteq
\doublehat{\Theta}_n \triangleq \left\{\theta \in \mathbb{R}^d\!:\!(\theta-\hat{\theta}_n)^\mathrm{T}V_n^\mathrm{T}H_n^{-1}V_n(\theta-\hat{\theta}_n)\leq r \right\},
\end{equation*}
where $r = q\hspace{0.2mm}$th largest value of $\gamma_i^*$, $i=1,\ldots,m-1$.

$\doublehat{\Theta}_n$ is the outer approximation we were looking for. Clearly it holds that
\begin{equation*}
\mathbb{P}\big(\theta^* \in \doublehat{\Theta}_n\big) \geq 1 - \frac{q}{m} = p,
\end{equation*}
for any finite $n$.
\fi
The pseudocode for computing $\doublehat{\Theta}_n$ is given in table \ref{ApproxTab}.

{\renewcommand{\arraystretch}{1.3}
\begin{table}[H]
\normalsize
\begin{center}
\begin{tabular}{|rlll|}
\hline
\multicolumn{4}{|c|}{\scshape Pseudocode: SPS-Outer-Approximation}\\
\hline \hline 1. & \multicolumn{3}{l|}{Compute the instrumental variables estimate} \\
&\multicolumn{3}{c|}{$\hat{\theta}_n=\bigg(\sum\limits_{t=1}^{n}\psi_t\varphi_t^\mathrm{T}\bigg)^{-1}\!\!\sum\limits_{t=1}^n\psi_t Y_t$;}\\
2. & \multicolumn{3}{l|}{For $i \in \{1, \dots, m-1\}$, solve the optimization} \\
& \multicolumn{3}{l|}{problem \eqref{OptDual}, and let $\gamma_i^*$ be the optimal value (or}\\
& \multicolumn{3}{l|}{$\infty$ if the problem is infeasible);}\\
3. & \multicolumn{3}{l|}{Let $r$ be the $q$\hspace{0.3mm}th largest $\gamma_i^*$ value;}\\
4. & \multicolumn{3}{l|}{The outer approximation of the SPS confidence}  \\ & \multicolumn{3}{l|}{region is given by the ellipsoid} \\
& \multicolumn{3}{l|}{$\doublehat{\Theta}_n =\big\{\theta \in \mathbb{R}^d\!:\!(\theta-\hat{\theta}_n)^{\mathrm{T}}V_n^\mathrm{T}H_n^{-1}V_n(\theta-\hat{\theta}_n)\leq r \big\}$.}\\
\hline
\end{tabular}
\end{center}
\caption{}\label{ApproxTab}
\vspace{-5mm}
\end{table}}
\section{Numerical experiments}\label{NumericalExp}
In this section we illustrate SPS with numerical experiments. Firstly, we apply the method to a simple first-order ARX system. Then, SPS is applied to a real-world identification problem, with the purpose of showing that the method is robust against the assumptions from which the guarantees provided in this paper are established.
\subsection{Simulation example}
We consider the following data generating ARX system
\[
Y_t = a^*Y_{t-1} + b^*U_{t} + N_t,
\]
where $a^*=0.7,b^*=1$, and $\{U_{t}\}$ is a sequence of random inputs generated as
\[
U_{t} = 0.75U_{t-1} + V_{t},
\]
being $\{V_{t}\}$ a sequence of i.i.d.\ Gaussian random variables $N(0,1)$. $\{N_t\}$ is a sequence of i.i.d.\ Laplacian random variables with zero mean and variance 1. We consider a finite sample of size $n$, that consists of couples $\{(Y_t, \varphi_t)\}_{t=1}^n$.

The instrumental variables $\{\psi_t\}_{t=1}^n$ are constructed from the data. In particular, we replace the autoregressive components of the regressors $\varphi_t$, for $t=2,\ldots,n$, with reconstructed outputs. Firstly we find an estimate $\hat{\theta}_{\mathrm{LS}}$ of the true parameter via least squares on $\{(Y_t, \varphi_t)\}_{t=1}^n$, and then we use such estimate\footnote{We could also use a {\em guess} (even imprecise) of the true parameter coming from some {\em a-priori} knowledge.} to build the noise-free sequence $\{\tilde{Y}_t\}_{t=1}^n$ using the following recursive procedure
\[
\tilde{Y_t} = \hat{a}\tilde{Y}_{t-1} + \hat{b}U_t,
\]
where $\hat{\theta}_{\mathrm{LS}}= [\hat{a}, \hat{b}]^\mathrm{T}$, and we use $Y_1$ as initialization value. Finally, the instrumental variables are
\[
{\psi}_t\triangleq[\tilde{Y}_{t-1}, U_t]^\mathrm{T}.
\]
Note that, rigorously speaking, these instrumental variables are not completely independent of the noise, due to the presence of the noise realization in the least squares estimate. However, in $\hat{\theta}_{\mathrm{LS}}$, the noise is {\em averaged out}, so that the effect of the noise is toned down. If the least squares estimate were built from a set independent of the one used by SPS then the constructed regions would be rigorous. Yet, the difference would be minimal, thus, for the sake of simplicity, we used just one data set.

\begin{figure}[H]
\includegraphics[width=8.5cm]{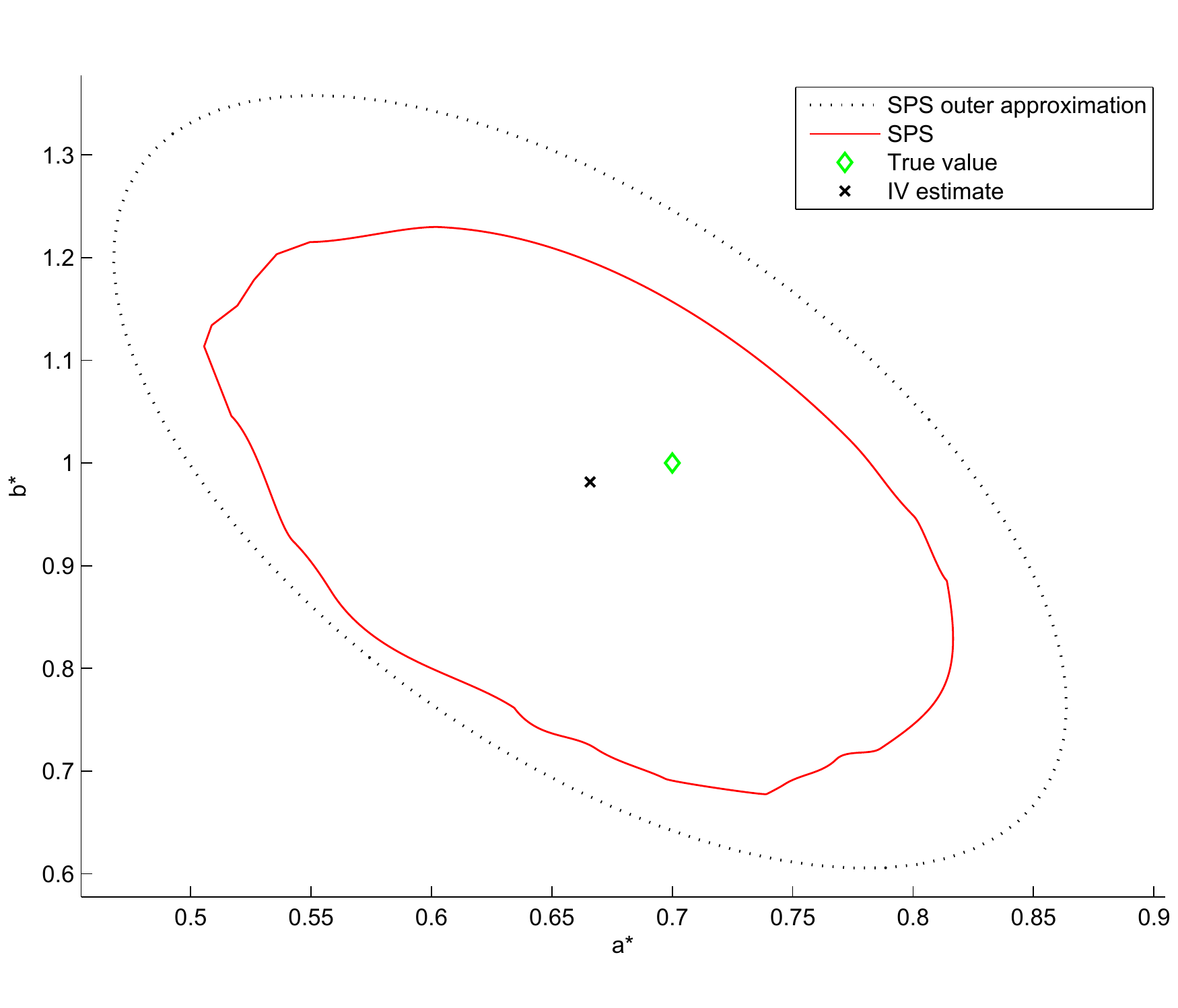}
\caption{$95\%$ confidence region, $n=25, m=100$.}
  \label{FigureExample}
\end{figure}

Based on $n=25$ data points $\{(Y_t, \varphi_t)\}_{t=1}^{25}$ we want to find a $95\%$ confidence region for $\theta^*$. We build 99 sign-perturbed sums ($m$ is set to 100), and the confidence region is constructed as the values of $\theta$ for which at least $q=5$ of the $\|S_i(\theta)\|^2$, $i=1,\ldots,99$, functions are ``larger''\footnote{According to the strict total order $\succ_\pi$, with a random permutation $\pi$.} than $\|S_0(\theta)\|^2$. An example of constructed confidence region is illustrated in figure \ref{FigureExample}. The solid red line has been obtained by evaluating the SPS-Indicator$(\theta)$ function in table \ref{IndTab} on a very fine grid.
\subsection{Real-world data experiment}
Working with real-world data is almost always a challenge. Usually, the user can only presume the nature of the {\em best} mathematical representation of the system, and most of the times the real system does not lie in the model class. Moreover, the knowledge on the noise characteristics is limited. All these issues make the identification process much more complicated. Nevertheless, we still want to apply SPS in such a scenario, and even though the theoretical results cannot be expected to hold rigorously, since, e.g., the real system does not lie in the model class, we hope that they hold approximately.

Our real-world data set comes from the photovoltaic energy production measurements of a prototype energy-positive public lighting microgrid (E+Grid) system \cite{BalazsPVPaper}. In particular, the available data contain the hourly historical progression of the amount of energy produced.

The model class is an ARX(5, 4), i.e.,
\[
Y_t = \sum_{i=1}^{5}a_iY_{t-i} + \sum_{i=1}^{4}b_iU_{t-i+1} + N_t= \varphi_t^\mathrm{T}\theta + N_t,
\]
where $Y_t$ is the amount of produced energy and $U_t$ is an auxiliary input given by the clear-sky predictions of the amount of energy produced (see \cite{BalazsPVPaper} for more details).

To carry out our tests, we first estimated via least squares a ``true parameter'' $\hat{\theta}^*$ based on the first half of the large (more than 4200 observations) data set available. After $\hat{\theta}^*\triangleq[\hat{a}^*, \hat{b}^*]^\mathrm{T}$ was found, the residuals $\varepsilon_t = Y_t - \sum_{i=1}^{5}\hat{a}^*_iY_{t-i} - \sum_{i=1}^{4}\hat{b}^*_iU_{t-i+1}$ were tested with the Durbin-Watson algorithm, \cite{durbin1950}, which returned a p-value bigger than 95\% for the uncorrelation hypothesis, supporting the choice of the orders 5 and 4 \cite{TesiVolpe}.

Then, SPS was used with the second half of the data set. The instrumental variables $\{\psi_t\}$ were built from the data by replacing the autoregressive components of the regressor with a reconstructed noise-independent trajectory of the output $\{\tilde{Y}_t\}$, similarly to what has been done in the previous example. The estimate of the ``true parameter'' used to build such a sequence was obtained via least squares on an extra subset of data consisting of 100 samples, which was not used later.
%
%The orders 5 and 4 were carefully selected via statistical tests (in particular, the Durbin-Watson test, \cite{durbin1950}, was used to verify that the residuals with standard identification procedure were uncorrelated).

Finally, we evaluated the empirical probability with which $\hat{\theta}^*$ belonged to the SPS regions that were built using many (1000) different data subsets, in a Monte Carlo approach. Each subset was constructed with pairs $\{(Y_t, \varphi_t)\}$ drawn randomly (non-sequentially) from the second half of the global data set. The size of each subset varied from 75 to 250 observations, and the parameter $m, q$ were always set, respectively, to 100 and 10, looking for a region of (desired) confidence probability equal to $90\%$.

The final results, illustrated in table \ref{TabResults}, show a good adherence between theory and empirical results.
\begin{table}[H]
\centering
\begin{tabular}{|c||c|}
\hline
$n$ & Empirical confidence  \\\hline\hline
75  &  0.886\\\hline
100  &  0.900\\\hline
150  &  0.886\\\hline
200  &  0.906\\\hline
250  &  0.910\\\hline
\end{tabular}
\caption{}
\label{TabResults}
\vspace{-5mm}
\end{table}
\section{Concluding remarks}
A new SPS algorithm has been proposed in this paper that, unlike the original version of SPS, can be used when the regressors contain past values of the system output, which makes it suitable for the identification of ARX systems. The algorithm makes use of instrumental variables (IV). However, it has to be noted that the reason for using an IV with SPS is quite different from other IV system identification methods. Particularly, in this version of SPS the IV does not counteract the presence of correlated noise, as it is in other IV approaches, and in fact the noise terms are supposed to form an independent pattern in this paper. Instead, the IV is introduced to ease the implementation of the method which is explained by noting that the IV only contains exogenous variables that are not affected by the system noise so that no noise sign perturbation is required in the IV when the sign-perturbed functions are constructed. Along an alternative approach, one may consider using the initial regressor $\varphi_t$ in place of the IV, which might give better shaped regions. However, this would require a more cumbersome implementation of the algorithm for the sign perturbation of the regressor, as it is done in \cite{SPSPaper1}. An evaluation of the pros and cons of these two approaches will be the subject of future investigations.
\iflong
\section*{Appendix A\\Proof of Theorem \ref{TheoremExact}: Exact Confidence}
We begin with a definition and some lemmas\footnote{For the proofs of the lemmas refer to \cite{SPSPaper2}.}.

\medskip
\begin{definition}
{\em Let $Z_1, \dots, Z_{k}$ be a finite collection of random variables and $\succ$ a strict total order. If for all permutations $i_1, \dots, i_{k}$ of indices $1,\dots, k$ we have
\begin{equation*}
\mathbb{P}(Z_{i_k} \succ Z_{i_{k-1}} \succ \dots \succ Z_{i_{1}}) = \frac{1}{k!},
\end{equation*}
then we call $\{Z_i\}$ uniformly ordered w.r.t.\ order $\succ$.}
\end{definition}
\smallskip
\begin{lemma}
\label{LemmaRandomSigns} {\em Let $\alpha, \beta_1, \dots, \beta_k$ be i.i.d.\ random signs,
then the random variables $\alpha, \alpha \cdot \beta_1, \dots, \alpha \cdot \beta_k$ are i.i.d.\ random signs.}
\end{lemma}
\smallskip
\begin{lemma}
\label{LemmaRealizations}
{\em Let $X$ and $Y$ be two independent, $\mathbb{R}^d$-valued and $\mathbb{R}^k$-valued random vectors, respectively. Let us consider a (measurable) function
$g: \mathbb{R}^d \times \mathbb{R}^k \to \mathbb{R}$ and a (measurable) set $A \subseteq \mathbb{R}$. If we have $\,\mathbb{P}(\,g(x,Y) \in A\,) = p$, for all (constant) $\,x \in \mathbb{R}^d$, then we also have $\,\mathbb{P}(\,g(X,Y) \in A\,) = p$.}
\end{lemma}
\smallskip
The following lemma highlights an important property of the $\succ_{\pi}$ relation that was introduced in Section \ref{SPSIV}.
\smallskip
\begin{lemma}
\label{LemmaIidCase}
{\em Let $Z_1, \dots, Z_{k}$ be real-valued, i.i.d. random variables. Then, they are uniformly ordered w.r.t.\ $\succ_{\pi}$.}
\end{lemma}
\subsection*{Proof of Theorem \ref{TheoremExact}}
By construction, parameter $\theta^*$ is in the confidence region if
$\mathcal{R}(\theta^*) \leq m-q$.
This means that  $\|S_0(\theta^*)\|^2$ takes one of the positions $1, \dots, m-q$ in the ascending order (w.r.t. $\succ_{\pi}$) of variables $\{\|S_i(\theta^*)\|^2\}$. We are going to prove that the $\{\|S_i(\theta^*)\|^2\}$ are {\em uniformly ordered}, hence $\|S_0(\theta^*)\|^2$ takes each position in the ordering with probability $1/m$, thus its rank is at most $m-q$ with probability $1-q/m$.

First, we fix a realization of the instrumental variables, by conditioning on the $\sigma$-algebra generated by them, and we will apply the following results realization-wise since noise and instrumental variables are independent by definition.

Note that for $\theta=\theta^*$, {\em all} $S_i(\cdot)$ functions have the form
\begin{equation*}
S_i(\theta^*) = H_n^{-\frac{1}{2}} \frac{1}{n} \sum_{t=1}^{n}{\, \alpha_{i,t} \, \psi_t N_t},
\end{equation*}
for all $i \in \{0, \dots, m-1\}$, where $\alpha_{0,t} \triangleq 1$, $t \in \{1, \dots, n\}$.

Therefore, all the $S_i(\cdot)$ functions depend on the perturbed noise sequence, $\{\alpha_{i,t}N_t\}$, via the {\em same} function for all $i$, which we denote by $S(\alpha_{i,1}N_{1}, \dots, \alpha_{i,n}N_n) \triangleq S_i(\theta^*)$.

Since each $N_t$ is symmetric, $\mathrm{sign}(N_t)$ and $|N_t|$ are independent. Then, for all $i$ and $t$, we introduce $\gamma_{i,t} \triangleq \alpha_{i,t}\, \mathrm{sign}(N_t)$, and since $\{\alpha_{i,t}\}$ are i.i.d.\ random signs, also $\gamma_{i,t}$ are i.i.d.\ random signs (Lemma \ref{LemmaRandomSigns}). Moreover, they are independent of $\{|N_t|\}$.

After fixing a {\em realization} of $\{|N_t|\}$, called $\{v_t\}$, we define the real-valued variables $\{Z_i\}$ by
\begin{equation*}
Z_i \,\triangleq\, \|S(\gamma_{i,1} v_1, \dots, \gamma_{i,n} v_n)\|^2\!\!.
\end{equation*}

Such $\{Z_i\}$ are i.i.d.\ random variables, and, in view of Lemma 3, they are uniformly ordered with respect to $\succ_{\pi}$.

So far we have proved the theorem assuming that the absolute values of the noises are constant, namely, the uniform ordering property was achieved by fixing a realization of $\{|N_t|\}$. However, the probabilities obtained are {\em independent of the particular realization} of $\{|N_t|\}$, hence, Lemma \ref{LemmaRealizations} can be applied to relax fixing the realization (i.e., in Lemma \ref{LemmaRealizations}, $X$ plays the role of $\{|N_t|\}$ and $Y$ incorporates the other random variables), and obtain the unconditional uniform ordering property of $\{\|S_i(\theta^*)\|^2\}$, from which the theorem follows. $\Box$
\section*{Appendix B\\Proof of Theorem \ref{TheoremStrongCons}: Strong Consistency}
We will prove that for any fixed (constant) $\theta' \neq \theta^*$, $\| S_0(\theta')\|^2\xrightarrow{\textrm{a.s.}}(\theta^* - \theta')^\textrm{T}V^\mathrm{T}H^{-1}V(\theta^* - \theta')$, which is larger than zero (using the strict positive definiteness of $H$, i.e., \ref{RPosDef}, and the invertibility of V, i.e., \ref{VInvertibility}), while, for $i \neq 0, \| S_i(\theta')\|^2\xrightarrow{\textrm{a.s.}}0$, as $n \rightarrow \infty$. This implies that, as $n$ grows, $\| S_0(\theta')\|^2$ will be ranked as the biggest element in the ordering, and therefore $\theta'$ will (almost surely) be excluded from the confidence region as $n \rightarrow \infty$. 
As done in the proof of Theorem \ref{TheoremExact}, we derive the results for a fixed  realization of the instrumental variables. Since instrumental variables and the noise $N_t$ are independent, the obtained results hold true on the whole probability space (almost surely).

Using the notation $\tilde\theta = \theta^* - \theta'$, $S_0(\theta')$ can be written as
\begin{align*}
S_0(\theta') &= H_n^{-\frac{1}{2}} \frac{1}{n}\sum_{t=1}^{n}\psi_t(Y_t - \varphi_t^\textrm{T}\theta')\\
&= H_n^{-\frac{1}{2}}\frac{1}{n}\sum_{t=1}^{n}\psi_t\varphi_t^\textrm{T}\tilde\theta + H_n^{-\frac{1}{2}}\frac{1}{n}\sum_{t=1}^{n}\psi_tN_t.
\end{align*}
The two terms will be analyzed separately. The convergence of the first term follows directly from \ref{RPosDef}, \ref{VInvertibility}, and by noticing that $(\cdot)^{\frac{1}{2}}$ is a continuous matrix function. Thus,
 \[
 H_n^{-\frac{1}{2}}\frac{1}{n}\sum_{t=1}^{n}\psi_t\varphi_t^\textrm{T}\tilde\theta=H_n^{-\frac{1}{2}}V_n\tilde\theta\xrightarrow{\textrm{a.s.}}H^{-\frac{1}{2}}V\tilde\theta,\textrm{ as } n\rightarrow\infty.
 \]
 The convergence of the second term will now be proved from the
 component-wise application of the Kolmogorov's strong law of large numbers (SLLN) for independent variables, \cite{Shiryaev1997}. Observe that $\{H_n^{-1/2}\}$ is a convergent sequence, just as $\{V_n\}$, so that for our purpose we only need to prove that the other part of the product
goes to zero (a.s.). By using the Cauchy-Schwarz inequality, \ref{InstrumentsGrowthRestriction}, and \ref{NoiseGrowthRestriction}, we have
\begin{eqnarray*}
\sum_{t=1}^{\infty}\frac{\mathbb{E}[\psi_{t,j}^2N_t^2]}{t^2}\leq\sum_{t=1}^{\infty}\frac{\| \psi_t\|^2}{t}\frac{\mathbb{E}[N_t^2]}{t}\leq\\
\sqrt{\sum_{t=1}^{\infty}\frac{\| \psi_t\|^4}{t^2}}\sqrt{\sum_{t=1}^{\infty}\frac{\mathbb{E}[N_t^2]^2}{t^2}}<\infty.
\end{eqnarray*}
Hence, the Kolmogorov's condition holds true and it holds that (SLLN)
\[
H_n^{-\frac{1}{2}}\frac{1}{n}\sum_{t=1}^{n}\psi_tN_t\xrightarrow{\textrm{a.s.}}0,\textrm{ as } n\rightarrow\infty.
\]
Using the two results we obtain
\[
\| S_0(\theta')\|^2\xrightarrow{\textrm{a.s.}}(\theta^* - \theta')^\textrm{T}V^\textrm{T}H^{-1}V(\theta^* - \theta') > 0,
\]
since $V$ is full rank, so that $V^\textrm{T}H^{-1}V$ is positive definite.

Now, we investigate the asymptotic behaviour of $S_i(\theta')$,
\begin{align*}
S_i(\theta') &= H_n^{-\frac{1}{2}} \frac{1}{n}\sum_{t=1}^{n}\alpha_{i,t}\psi_t(Y_t - \varphi_t^\textrm{T}\theta') \\
&= H_n^{-\frac{1}{2}}\frac{1}{n}\sum_{t=1}^{n}\alpha_{i,t}\psi_t\varphi_t^\textrm{T}\tilde\theta + H_n^{-\frac{1}{2}}\frac{1}{n}\sum_{t=1}^{n}\alpha_{i,t}\psi_tN_t.
\end{align*}
Again, we will inspect the asymptotic behaviour of the two
terms separately. The convergence of the second term follows
immediately from our previous argument, since the variance
of $\alpha_{i,t}\psi_tN_t$ is the same as the variance of $\psi_tN_t$. Thus,
\[
H_n^{-\frac{1}{2}}\frac{1}{n}\sum_{t=1}^{n}\alpha_{i,t}\psi_tN_t\xrightarrow{\textrm{a.s.}}0,\textrm{ as } n\rightarrow\infty.
\]
\addtolength{\textheight}{-5cm}
For the first term, since $\{H_n^{-\frac{1}{2}}\}$ is convergent and $\tilde\theta$ is constant, it is enough to show that $\frac{1}{n}\sum_{t=1}^{n}\alpha_{i,t}[\psi_t\varphi_t^\textrm{T}]_{j, k}$ converges almost surely to 0 for each $j$ and $k$. In order to do that, we fix a realization of the noise, so that $\{\alpha_{i,t}[\psi_t\varphi_t^\textrm{T}]_{j, k}\}$ becomes a sequence of (conditionally) independent random variables with (conditional) covariances $[\psi_t\varphi_t^\textrm{T}]^2_{j, k}$. From \ref{RegressorGrowthRestriction} and \ref{InstrumentsGrowthRestriction},
\begin{eqnarray*}
\sum_{t=1}^{\infty}\frac{[\psi_t\varphi_t^\textrm{T}]^2_{j, k}}{t^2} = \sum_{t=1}^{\infty}\frac{\psi_{t,j}^2\varphi_{t,k}^2}{t^2} \leq\\ \sqrt{\sum_{t=1}^{\infty}\frac{\| \psi_t\|^4}{t^2}}\sqrt{\sum_{t=1}^{\infty}\frac{\| \varphi_t\|^4}{t^2}} < \infty.
\end{eqnarray*}
Therefore, using the SLLN
\[
H_n^{-\frac{1}{2}}\frac{1}{n}\sum_{t=1}^{n}\alpha_{i,t}\psi_t\varphi_t^\textrm{T}\tilde\theta\xrightarrow{\textrm{a.s.}}0,\textrm{ as } n\rightarrow\infty
\]
holds true for (almost) any noise realization, and therefore holds true unconditionally.

Now, since $\|S_0(\theta')\|^2\xrightarrow{\textrm{a.s.}}(\theta^* - \theta')^\textrm{T}V^\textrm{T}H^{-1}V(\theta^* - \theta')$ and $\| S_i(\theta')\|^2\xrightarrow{\textrm{a.s.}}0, i \neq 0$, we show that eventually the
confidence region will (a.s.) be contained in a ball of radius $\varepsilon$  around the true parameter, $\theta^*$, for any positive $\varepsilon$.

From the previous results, we know that the event that for each $i \in \{0,\ldots,m-1\}$ the functions $\|S_i(\theta')\|^2$ converge is a set of probability $1$. Fix an outcome from this set, and define

%
%Let $(\Omega, \mathcal{F}, \mathbb{P})$ denote the background probability space, where $\Omega$ is the sample space, $\mathcal{F}$ is the $\sigma$-algebra of events (conditioned to the realization of the instrumental variables), and $\mathbb{P}$ is the probability measure. Then, there is an event $F_0 \in \mathcal{F}$ such that $\mathbb{P}(F_0)=1$  and for all $\omega \in F_0$, for each $j \in \{0,\ldots,m-1\}$, the functions $\|S_i(\theta')\|^2$ converge.

\begin{align*}
\Gamma_{i,n} &\triangleq\frac{1}{n}\sum_{t=1}^{n}\alpha_{i,t}\psi_t\varphi_t^\textrm{T},
\\
\gamma_{i,n}&\triangleq \frac{1}{n}\sum_{t=1}^{n}\alpha_{i,t}\psi_tN_t,
\\
\upsilon_n&\triangleq \frac{1}{n}\sum_{t=1}^{n}\psi_tN_t.
\end{align*}
Given the previous results, for each $\delta >0$ there must be an $N > 0$ such that for $n \geq N$ (for all $i \neq 0$),
\begin{align*}
\| H_n^{-\frac{1}{2}}V_n - H^{-\frac{1}{2}}V\|&\leq \delta, & \| H_n^{-\frac{1}{2}}\upsilon_n\|&\leq \delta,\\\| H_n^{-\frac{1}{2}}\Gamma_{i,n}\|&\leq \delta, & \| H_n^{-\frac{1}{2}}\gamma_{i,n} \|&\leq \delta,
\end{align*}
where $\| \cdot \|$ indicates the spectral
norm (if its argument is a matrix), i.e.\ the matrix norm induced by the Euclidean vector norm. Take $n\geq N$, then
\begin{align*}
\| S_0(\theta')\| \!&= \!\| H_n^{-\frac{1}{2}}V_n\tilde{\theta} + H_n^{-\frac{1}{2}}\upsilon_n\|
\\
&=\!\| (H_n^{-\frac{1}{2}}V_n \!-\! H^{-\frac{1}{2}}\!V)\tilde{\theta} \!+\! H^{-\frac{1}{2}}V\tilde{\theta} \!+\! H_n^{-\frac{1}{2}}\!\upsilon_n\|
\\
&\geq\!\lambda_{\textrm{min}}(H^{-\frac{1}{2}}V)\|\tilde{\theta}\| - \delta\|\tilde{\theta}\| - \delta,
\end{align*}
where $\lambda_\text{min}(\cdot)$ denotes the smallest eigenvalue. On the other hand, we also have
\begin{align*}
\| S_i(\theta')\| &\!=\!\| H_n^{\!-\frac{1}{2}}\Gamma_{i,n}\tilde{\theta} + H_n^{\!-\frac{1}{2}}\!\gamma_{i,n}\|
\\
&\!\leq\!\| H_n^{\!-\frac{1}{2}}	\Gamma_{i,n} \|\|\tilde{\theta}\| \!+\! \| H_n^{\!-\frac{1}{2}}\!\gamma_{i,n}\|\!\leq\!\delta\|\tilde{\theta}\| \!+\! \delta.
\end{align*}
We have $\| S_i(\theta') \|<\| S_0(\theta')\|$ for all $\theta'$ that satisfy
\[
\delta\|\tilde{\theta}\| + \delta<\lambda_{\text{min}}(H^{-\frac{1}{2}}V)\|\tilde{\theta}\| - \delta\|\tilde{\theta}\| - \delta,
\]
which can be rewritten as
\[
\kappa_0(\delta)\triangleq\frac{2\delta}{\lambda_{\text{min}}(H^{-\frac{1}{2}}V) - 2\delta}<\|\tilde{\theta}\|,
\]
therefore, those $\theta'$ for which $\kappa_0(\delta)<\|\theta^* - \theta'\|$ are not included in the confidence region $\widehat{\Theta}_n$, for $n\geq N$.  Finally, by setting $\delta := (\varepsilon\lambda_{\text{min}}(H^{-\frac{1}{2}}V))/(2+ 2\varepsilon)$ we can prove the statement of the theorem for any positive $\varepsilon$. $\Box$
\else
\addtolength{\textheight}{-11cm}
\fi
%%%%%%%%%%%%%%%%%%%%%%%%%%%%%%%%%%%%%%%%%%%%%%%%%%%%%%%%%%%%%%%%%%%%%%%%%%%%%%%%

%%%%%%%%%%%%%%%%%%%%%%%%%%%%%%%%%%%%%%%%%%%%%%%%%%%%%%%%%%%%%%%%%%%%%%%%%%%%%%%%

\bibliography{Bibl}{}
\bibliographystyle{plain}
\end{document}